\documentclass[runningheads]{llncs}
\usepackage[T1]{fontenc}
\usepackage{amsmath,amssymb} %,amsthm}
\usepackage{graphicx}
\usepackage{hyperref}
\usepackage{xspace}
\usepackage{wrapfig}
\usepackage{multirow}

\usepackage{color}

\usepackage{scalerel}

\newcounter{exampcount}
{\hfill$\blacksquare$\vskip6pt}

\newcommand{\startpara}[1]{{%
\vskip6pt\noindent
{\bf #1.}}}

\newcommand{\coalition}[1]{\langle \! \langle {#1} \rangle \! \rangle}

\def\future{{\mathtt{F}\ }}

\newcommand\probopP{{\mathtt P}}
\newcommand\nashop[3]{\coalition{#1}_{#2}(#3)}

\newcommand\probop[2]{\probopP_{#1}[\,{#2}\,]}

\newcommand{\pctlstar}[1]{PCTL* }
\newcommand{\rpatlstar}[1]{RPATL* }

\newcommand{\coal}[1]{\lla #1 \rra}
\newcommand{\lla}{\langle\!\langle}
\newcommand{\rra}{\rangle\!\rangle}

\title{Multi-Agent Verification and Control \\ with Probabilistic Model Checking}

\author{David Parker}

\institute{Department of Computer Science, University of Oxford, Oxford, UK \email{david.parker@cs.ox.ac.uk}}

\begin{document}

\maketitle

\begin{abstract}
Probabilistic model checking is a technique for
formal automated reasoning about software or hardware
systems that operate in the context of uncertainty or stochasticity.
It builds upon ideas and techniques from a diverse range of fields,
from logic, automata and graph theory,
to optimisation, numerical methods and control.
In recent years, probabilistic model checking has also been extended
to integrate ideas from game theory,
notably using models such as stochastic games and solution concepts such as equilibria,
to formally verify the interaction of multiple rational agents with distinct objectives.
This provides a means to reason flexibly about
agents acting in either an adversarial or a collaborative fashion,
and opens up opportunities to tackle new problems within, for example,
artificial intelligence, robotics and autonomous systems.
In this paper, we summarise some of the advances in this area,
and highlight applications for which they have already been used.
We discuss how the strengths of probabilistic model checking
apply, or have the potential to apply, to the multi-agent setting
and outline some of the key challenges required to make further progress in this field.
\end{abstract}

%============================================================

\section{Introduction}\label{sec:intro}

Probabilistic model checking is a fully automated approach
for formal reasoning about systems exhibiting uncertain behaviour,
arising for example due to faulty hardware, unpredictable operating environments or the use of randomisation.
Probabilistic models, such as Markov chains, Markov decision processes (MDPs) or their variants,
are systematically explored and analysed in order to establish whether
formal specifications given in temporal logic are satisfied.
For models such as MDPs, controllers (policies) can be automatically
generated to ensure that such specifications are met, or are optimised for.

Probabilistic model checking builds on concepts and techniques from a wide array of other fields.
Its roots lie in formal verification, and it relies heavily on the use of logic and automata.
Since models typically need to be solved or optimised numerically,
it also adopts methods from Markov chains, control theory, optimisation and, increasingly, from various areas of artificial intelligence.

In the this paper, we discuss the integration of probabilistic model checking
with ideas from \emph{game theory}, facilitating the verification or control of multi-agent systems in the context of uncertainty.
We focus on \emph{stochastic games}, which model the interaction between multiple agents (players)
operating in a dynamic, stochastic environment.
They were introduced in the 1950s by Shapley~\cite{Sha53},
generalising the classic model of Markov decision processes~\cite{Bel57} to the case of multiple players,
and techniques for their solution have been well studied~\cite{FV96}.

In the context of formal verification,
game-theoretic modelling has a number of natural applications,
in particular when an agent interacts with an \emph{adversary} that has opposing goals,
for example a defender and an attacker in the context of a computer security protocol
or honest and malicious participants in a distributed consensus algorithm.
For verification or control problems in stochastic environments,
game-based models also underlie methods for \emph{robust} verification,
using worst-case assumptions of epistemic model uncertainty.
Furthermore, game theory provides tools for controller synthesis
in a more \emph{cooperative} setting, for example via the use of
\emph{equilibria} representing strategies for collections of players
with differing, but not strictly opposing objectives.

Verification problems for stochastic games have been quite extensively studied (see, e.g.,~\cite{Cha07b,Umm10,CH12})
and in recent years, probabilistic model checking frameworks and tools
have been developed (e.g.,~\cite{CFK+13b,KNPS21,KNPS20})
and applied to a variety of problem domains.
In the next section, we summarise some of these advances.
We then go on to discuss the particular strengths of probabilistic model checking,
the ways in which these are applicable to multi-agent models
and some of the remaining challenges that exist in the field.

%============================================================

\section{Model Checking for Stochastic Games}\label{sec:mcsg}

\subsection{Turn-based stochastic games}

Stochastic games comprise a set of $n$ players
making a sequence of decisions that have stochastic outcomes.
The way in which players interact can be modelled in various ways.
The simplest is with a \emph{turn-based stochastic game} (TSG).
The state space $S$ of the game is partitioned into disjoint sets
$S_1\uplus\dots\uplus S_n=S$, where states in $S_i$ are controlled by player $i$.
Players choose between actions from a set $A$
(for simplicity, let us assume that all actions are available to be taken in each state)
and the dynamics of the model is captured, like for an MDP,
by a probabilistic transition function $P:S\times A\times S\rightarrow[0,1]$,
with $P(s,a,s')$ giving the probability to move to state $s'$ when action $a$ is taken in state $s$.

A probabilistic model checking framework for TSGs is presented in~\cite{CFK+13b},
which proposes the logic rPATL (and its generalisation rPATL*)
for specifying \emph{zero-sum} properties of stochastic games,
adding probabilistic and reward operators to the well known game logic
ATL (alternating temporal logic)~\cite{alur2002alternating}.
A simple (numerical) query is 
$\coal{\mathit{ag}_1,\mathit{ag}_2}\probop{\max=?}{\future\mathsf{goal}}$,
which asks for the maximum probability of reaching a set of states $\mathsf{goal}\subseteq S$
that is achievable by a coalition of the players $\mathit{ag}_1$ and $\mathit{ag}_2$,
assuming that any other players in the game have the directly opposing objective
of minimising the probability of this event.

Despite a worse time complexity than MDPs for the core underlying problems
(e.g., computing optimal reachability probabilities, for the query above, is in NP$\,\cap\,$co-NP, rather than PTIME),
\cite{CFK+13b} shows that value iteration (dynamic programming)
is in practice an effective and scalable approach.
For maximising probabilistic reachability with two players,
the values $x^k_s$ defined below converge, as $k\rightarrow\infty$, to the required probability for each state $s$:
$$x^k_s =
\begin{cases}
1 & s\in\mathsf{goal} \\
0 & s\not\in\mathsf{goal}\mbox{ and }k=0\\
\max_{a\in A}\sum_{s'\in S}P(s,a,s')\cdot x^{k-1}_{s'} & s\in S_1\backslash\mathsf{goal}\mbox{ and }k>0\\
\min_{a\in A}\sum_{s'\in S}P(s,a,s')\cdot x^{k-1}_{s'} & s\in S_2\backslash\mathsf{goal}\mbox{ and }k>0\\
\end{cases}$$
The computation yields optimal strategies for players,
which are deterministic (i.e., pick a single action in each state)
and memoryless (i.e., do so regardless of history).
The model checking algorithm~\cite{CFK+13b} extends to
many other temporal operators including a variety of reward (or cost) based measures.

Subsequently, various other aspects of TSG model checking have been explored, including
the performance of different game solution techniques~\cite{KRSW22},
the use of interval iteration methods to improve accuracy and convergence~\cite{EKKW22},
trade-offs between multiple objectives~\cite{BKW17,CKWW20}
and the development of symbolic (decision diagram based) implementations to improve scalability~\cite{KNPS22b}.

Despite the relative simplicity of TSGs as a modelling formalism,
they have been shown to be appropriate for various scenarios
in which there is natural turn-based alternation between agents;
examples include human-robot control systems~\cite{FWH+15,JJK+18}
and self-adaptive software systems interacting with their environment~\cite{CGSP15}.

%-----------------------------------

\subsection{Concurrent stochastic games}

To provide a more realistic model of the concurrent execution of agents,
we can move to the more classic view of player interaction in stochastic games,
where players make their decisions simultaneously and independently.
To highlight the distinction with the turn-based model variant discussed above,
we call these \emph{concurrent stochastic games} (CSGs);
the same model is referred to elsewhere as, for example,
Markov games or multi-agent Markov decision processes.
In a CSG, each player $i$ has a separate set of actions $A_i$
and the probabilistic transition function $P:S\times (A_1\times\dots\times A_n) \times S\rightarrow[0,1]$
now models the resulting stochastic state update that arises for each possible joint player action.

Probabilistic model checking of CSGs against zero-sum objectives,
again using the logic rPATL, is proposed in~\cite{KNPS21}.
Crucially, optimal strategies for players are now randomised,
i.e., can choose a probability of selecting each action in each state,
however, a value iteration approach can again be adopted.
Consider again maximal probabilistic reachability for two players:
instead of simply picking the highest value action for one player in each state,
a one-shot matrix game $Z$, indexed over action sets $A_1$ and $A_2$, is solved at each state:
$$\begin{array}{ll}
x^k_s =
\begin{cases}
1 & s\in\mathsf{goal} \\
0 & s\not\in\mathsf{goal}\mbox{ and }k=0\\
\mathrm{val}(Z) & s\not\in\mathsf{goal}\mbox{ and }k>0 \\
\end{cases} &
\mbox{\hspace*{0.4cm}where } Z_{a,b}=\sum\limits_{s'\in S}P(s,(a,b),s')\cdot x^{k-1}_{s'}
\end{array}
$$
The matrix game $Z$ contains payoff values for player 1
and the value $\mathrm{val}(Z)$ of $Z$ is the optimal achievable value when player 2 minimises.
This is solved via a (small) linear programming problem over variables $\{p_{a}\,|\,a\in A_1\}$
which yields the optimal probabilities $p_{a}$ for player 1 to pick each action $a$.
Despite the increase in solution complexity with respect to TSGs,
results in \cite{KNPS21} show the feasibility of building and solving large CSGs
that model examples taken from robotics, computer security and communication protocols.
These also highlight deficiencies when the same examples are modelled with TSGs.

%-----------------------------------

\subsection{Equilibria for stochastic games}

Zero-sum objectives, for example specified in rPATL,
allow synthesis of optimal controllers in the context
of both stochasticity and adversarial behaviour.
But there are many instances where
agents do not have directly opposing goals.
The CSG probabilistic model checking framework has been extended
to incorporate non-zero-sum objectives such as Nash equilibria (NE)~\cite{KNPS21}.
Informally, strategies for a set of players with distinct, individual objectives
form an NE when there is no benefit to any of them
of unilaterally changing their strategy.

It is shown in ~\cite{KNPS21}, that by focusing on \emph{social welfare} NE,
which also maximise the sum of all players' objectives,
value iteration can again be applied.
Extending rPATL, we can write for example
$\nashop{\mathit{ag}_1{:}\mathit{ag}_2}{\max=?}{\probop{}{\future
\mathsf{goal}_1}{+}\probop{}{\future\mathsf{goal}_2}}$
to ask for the social welfare NE in which
two players maximise the probabilities of reaching distinct sets of state $\mathsf{goal}_1$ and $\mathsf{goal}_2$.
% The computation at each step of
Value iteration becomes:
$$\begin{array}{l}
x^k_s =
\begin{cases}
(1,1) & s\in\mathsf{goal}_1 \cap \mathsf{goal}_2 \\
(1,P^{\max}_{s,\mathsf{goal}_2}) & s\in\mathsf{goal}_1 \backslash \mathsf{goal}_2 \\
(P^{\max}_{s,\mathsf{goal}_1},1) & s\in\mathsf{goal}_2 \backslash \mathsf{goal}_1 \\
(0,0) & s\not\in(\mathsf{goal}_1 \!\cup \mathsf{goal}_2)\mbox{ and }k=0 \\
\mathrm{val}(Z^1,Z^2) & s\not\in(\mathsf{goal}_1 \!\cup \mathsf{goal}_2)\mbox{ and }k>0 \\
\end{cases}\\\\[-0.4em]
\mbox{\hspace*{0.2cm}where } Z^i_{a,b}=\sum\limits_{s'\in S}P(s,(a,b),s')\cdot x^{k-1}_{s'}(i),
\end{array}
$$
$\mathrm{val}(Z^1,Z^2)$ is the value of a \emph{bimatrix game}
and $P^{\max}_{s,\mathsf{goal}_i}$ is the maximum probability
of reaching $\mathsf{goal}_i$ from state $s$,
which can be computed independently by treating the stochastic game as an MDP.
The value of the (one-shot) game defined by payoff matrices $Z^i$ for player $i$
is a (social welfare) NE, computed in~\cite{KNPS21}
using an approach called labelled polytopes~\cite{LH64} and a reduction to SMT.
Optimal strategies (in fact, $\epsilon$-optimal strategies) can be extracted.
They are, as for zero-sum CSGs, randomised but also require memory.
% stopping games only

The move towards concurrent decision making over distinct objectives
opens up a variety of interesting directions for exploration.
Equilibria-based model checking of CSG is extended in several ways in~\cite{KNPS22}.
Firstly, \emph{correlated} equilibria allow players to coordinate
through the use of a (probabilistic) public signal,
which then dictates their individual strategies.
These are shown to be cheaper to compute
and potentially more equitable in the sense that they improve joint outcomes.
Secondly, \emph{social fairness} is presented as an alternative
optimality criterion to social welfare,
which minimises the \emph{difference} between players' objectives,
something that is ignored by by the latter criterion.

%============================================================

\section{Opportunities and Challenges}\label{sec:discuss}

Probabilistic model checking is a flexible technique,
which already applies to many different types of stochastic models
and temporal logic specifications.
On the one hand, the thread of research described above represents
a further evolution of the approach towards a new class of models and solution methods.
On the other, it represents an opportunity to apply the strengths of probabilistic model checking
to a variety of problem domains in which multi-agent approaches to modelling are applicable
and where guarantees on safety or reliability may be essential;
examples include multi-robot coordination, warehouse logistics, autonomous vehicles or robotic swarms.
We now discuss some of the key benefits of probabilistic model checking
and their relevance to the multi-agent setting.
We also summarise some of the challenges that arise as a result.

\startpara{Temporal logic}
Key to the success of model checking based techniques is the
ability to precisely and unambiguously specify desired system properties
in a formal language that is expressive enough to be practically useful,
but constrained enough that verification or control problems are practical.

Temporal logics such as rPATL and its extensions show that
it is feasible to combine quantitative aspects (probability, costs, rewards)
with reasoning about the strategies and objectives of multiple agents,
for both zero-sum optimality and equilibria of various types.
This combines naturally with the specification of temporal behaviour
using logics such as LTL, from simple reachability or reach-avoid goals,
to more complex sequences of events and long-run specifications.
The latter have been of increasing interest in, for example,
task specification in robotics~\cite{KGFP09} or reinforcement learning~\cite{HAGW21}.
Another key benefit here is the continual advances in translations from 
such logics to automata, facilitating algorithmic analysis.
From a multi-agent perspective, specifically, challenges include
more expressive reasoning about dependencies between strategies
or epistemic aspects, where logic extensions exist but model checking is challenging.

\startpara{Tool support and modelling languages}
The main functionality for model checking of stochastic games described in Section~\ref{sec:mcsg}
is implemented within the PRISM-games tool~\cite{KNPS20},
which has been developed over the past 10 years.
However, interest in verification for this class of models is growing and,
for the simpler model variant of TSGs, support now exists in multiple
actively developed probabilistic model checking tools \cite{PKPB21,Meg22,KRSW22,FHLSSTZ22}. % TEMPEST, PET, PRISM-games ext, EPMC

Currently, these tools share a common formalism for specifying models, % (spec)
namely PRISM-games's extension to stochastic games of the widely used PRISM modelling language.
Although relatively simple from a programming language perspective,
this has proved to be expressive enough for modelling across a large range of application domains.
Key modelling features include the ability to describe systems as the parallel composition
of multiple, interacting (sometimes duplicated) components
and the use of parameters for easy reconfiguration of models.
It also provides a common language for many different types of probabilistic models
through various features that can be combined, e.g.,
clocks (for \emph{real-time} modelling),
observations (for \emph{partially observable} models)
and model uncertainty such as transition probability intervals (for \emph{epistemic uncertainty}).

Component-based modelling is of particular benefit for the multi-agent setting,
but challenges remain as the modelling language evolves.
Examples include dealing with the subtleties that arise regarding how components
communicate and synchronise, particularly under partial observability,
and the specification of particular strategies for some agents.

\startpara{Exhaustive analysis}
Traditionally, a strength of model checking based techniques
is their focus on an exhaustive model analysis,
allowing them to identify (or prove the absence of)
corner cases representing erroneous or anomalous behaviour.
In the stochastic setting this remains true,
in particular for models that combine probabilistic and nondeterministic behaviour.
The subtle interaction between these aspects can be difficult to reason about without automated tools,
and exhaustive approaches can in these cases be preferable
to more approximate model solution methods, such as those based on simulation.

Adding multiple players only strengthens the case for these techniques.
For example, \cite{CFK+13b} identifies weaknesses in a distributed, randomised energy management protocol
arising when some participants behave selfishly; a simple incentive scheme is then shown to help alleviate this issue.
Multi-agent models allow a combination of \emph{control} and \emph{verification},
for example synthesising a controller (strategy) for one player, or coalition of players,
which can be verified to perform robustly in the context of adversarial behaviour by other players.

A natural direction is to then deploy verification to controllers
synthesised by more widely applicable and more scalable approaches
such as multi-agent reinforcement learning~\cite{ACS23}.
This brings challenges in terms of, for example,
extending probabilistic model checking to continuous state spaces,
and tighter integration with machine learning methods.
Progress in this direction includes the extension of CSGs to a \emph{neuro-symbolic} setting~\cite{YSD+22,nscsgs},
which incorporates neural networks for specific model components, e.g. for perception tasks.

\startpara{Further challenges}
In addition to those highlighted above, numerous other challenges exist in the field.
One perennial issue with model checking approaches is their \emph{scalability} to large systems.
Symbolic approaches, e.g., based on decision diagrams, have proved to be valuable
for probabilistic model checking, and also extended to TSGs~\cite{KNPS22b}.
However it is unclear how to adapt these to CSGs:
while value iteration is often amenable to a symbolic implementation,
methods such as linear programming or bimatrix game solution are typically not.

The use of modelling formalisms like the PRISM-games language
should also encourage the development of \emph{compositional} analysis techniques,
such as counter abstraction \cite{LP22} or assume-guarantee methods,
but progress remains limited in these directions within probabilistic model checking.
On a related note, while human-created models naturally exhibit high-level structure,
strategies synthesised by model checking tools typically do not.
This hinders comprehension and explainability,
which becomes more important when strategies, as here,
are more complex due to randomisation, memory and distribution across agents.

There are also major \emph{algorithmic} challenges which arise as the techniques
are applied to new problems. For example, there have been steady advances in
verification techniques for \emph{partially observable} MDPs,
but much less work on this topic for stochastic games.
Finally, there are potential benefits from further exploration of ideas from game theory,
e.g., other equilibria, such as Stackelberg (with applications, for instance, to security or automotive settings)
or the inclusion of epistemic aspects into logics and model checking algorithms.

%============================================================

\startpara{Acknowledgements}
This work was funded in part by the ERC under the European Union’s Horizon 2020 research and innovation programme (\href{http://www.fun2model.org}{FUN2MODEL}, grant agreement No.~834115).

\bibliographystyle{splncs04}
\bibliography{bib}

\end{document}